\documentclass[3p]{elsarticle}
\usepackage{amssymb}
\usepackage{bbm}
\usepackage{graphicx}
\usepackage{dcolumn}
\usepackage{bm}
\usepackage{epsfig}
\usepackage{subcaption}
\usepackage{multirow}
\usepackage{amsmath} 
\usepackage{hyperref}
\journal{Nuclear Physics B}
\usepackage{natbib}
\begin{document}

\begin{frontmatter}

\title{Neutrino Mass Matrices with Generalized CP Symmetries and Texture Zeros}

\author[1]{Sanjeev Kumar}
\ead{skverma@physics.du.ac.in}

\author[2]{Radha Raman Gautam}
\ead{gautamrrg@gmail.com}

\address[1]{Department of Physics and Astrophysics, University of Delhi, Delhi-110007, India}

\address[2]{Department of Physics, LBS Govt. Degree College, Saraswati Nagar, Shimla-171206, India}

\begin{abstract}
We investigate the properties of neutrino mass matrices that incorporate texture zeros and generalized CP symmetries associated with tribimaximal mixing. By combining these approaches, we derive predictive neutrino mass matrices and explore their implications for mass hierarchies, mixing angles, and CP-violating phases. We find that the three angles defining the generalized CP symmetries have narrow allowed ranges. We also obtain distinct correlations between the three mixing angles and the CP-violating phases that distinguish the various texture patterns from one another. Moreover, we compute the effective neutrino mass for neutrinoless double beta decay and the sum of neutrino masses. Our results highlight the predictability and testability of neutrino mass matrices with generalized CP symmetry.
\end{abstract}
\begin{keyword}
 Neutrino mass matrix, Generalized CP symmetry
\end{keyword}
\end{frontmatter}

\section{Introduction}
The discovery of neutrino oscillations indicates that neutrinos have non-zero masses and mix with one another. After this discovery, the precision measurements of lepton masses and mixings and determination of mass hierarchy and  CP violation in the lepton sector are the main focus of forthcoming neutrino experiments. These discoveries can help us in understanding of leptogenesis and the matter-antimatter asymmetry of the universe.  

The neutrino mixing matrix $U$, also known as the PMNS matrix, can be written as
\begin{small}
\begin{equation} \label{eqpdg}
U = \begin{pmatrix}
\cos \theta_{12}  \cos \theta_{13} & \sin \theta_{12} \cos \theta_{13}  & \sin \theta_{13} e^{-i \delta} \\
- \cos \theta_{23} \sin \theta_{12} - \cos \theta_{12} \sin \theta_{13} \sin \theta_{23} e^{i \delta} &
\cos \theta_{12} \cos \theta_{23} - \sin \theta_{12} \sin \theta_{13} \sin \theta_{23} e^{i \delta} & \cos \theta_{13} \sin \theta_{23} \\
\sin \theta_{12} \sin \theta_{23} - \cos \theta_{12} \sin \theta_{13} \cos \theta_{23} e^{i \delta} &
-\cos \theta_{12} \sin \theta_{23} - \sin \theta_{12} \sin \theta_{13} \cos \theta_{23} e^{i \delta} & \cos \theta_{13} \cos \theta_{23} \\
\end{pmatrix}
P
\end{equation}
\end{small}
in its PDG parametrization \cite{ParticleDataGroup:2018ovx} after a slight variation in the definition of the phase matrix $P$:
\begin{equation}
P =\begin{pmatrix}
    e^{-i \phi_1}      &  0 & 0 \\
   0    &  e^{-i \phi_2}  & 0 \\
   0      & 0 &  e^{-i \phi_3}
\end{pmatrix}.
\end{equation}
The mixing matrix $U$ is parameterized by three mixing angles ($\theta_{12}$, $\theta_{23}$,  and $\theta_{13}$), one Dirac type CP-violating phase ($\delta$), and two Majorana type CP-violating phases ($\phi_{12} = \phi_{2}-\phi_1$ and $\phi_{13}=\phi_3-\phi_1$). 

The lepton mixing matrix $U$ parameterizes the mixing between the three flavors of neutrinos and diagonalizes the neutrino mass matrix $M_{\nu}$ as 
\begin{equation}
U^T M_{\nu} U = \text{diag} (m_1,m_2,m_3). 
\end{equation}
The matrix $M_{\nu}$ can be parameterized as a function of three mixing angles, three neutrino masses, and three CP-violating phases. However, neutrino oscillation experiments are sensitive only to the  mass-squared differences ($\Delta m^2_{21}$ and $\Delta m^2_{31}$, where $\Delta m_{ij}^2 = m_i^2 - m_j^2$, $i,j = 1,2,3$). We know definitely that the second neutrino mass eigenstate is heavier than the first mass eigenstate ($m_2>m_1$). But, we cannot say so about the third mass eigenstate as the sign of  $\Delta m^2_{31}$ is still undetermined. The current experiments allow both normal hierarchy (NH) with $m_1<m_3$ and inverted hierarchy (IH) with $m_1>m_3$ corresponding to the two sign choices of  $\Delta m^2_{31}$. 

The neutrino oscillation experiments provide information only about the neutrino mass-squared differences and not on the absolute neutrino mass scale. The constraints on the absolute neutrino mass come from the experiments on beta decay, neutrinoless double beta decay, and cosmological experiments.  The current understanding of these neutrino experiments has been reviewed in Ref. \cite{SajjadAthar:2021prg}.
 
There exist certain special patterns in the mixing matrix $U$ that can be attributed to a residual flavor symmetry of $M_{\nu}$. Such residual symmetries arise when $M_{\nu}$ is invariant under certain non-trivial transformations $G_i$ of the form 
\begin{equation}
G_i= U d_i U^{\dagger}
\end{equation}
where 
\begin{equation}
d_1 = \text{diag}(1,-1,-1),
\end{equation}
\begin{equation}
d_2 = \text{diag}(-1,1,-1),
\end{equation}
and 
\begin{equation}
d_3 = \text{diag}(-1,-1,1).
\end{equation}
In other words, 
\begin{equation}
G_i^T M_{\nu} G_i=M_{\nu} \text{ for } i=1,2,3. 
\end{equation}
The concept of residual flavor symmetries has been thoroughly explored in the literature \cite{Altarelli:2010gt, Ishimori:2010au, Grimus:2011fk, King:2013eh, King:2014nza}. One of the most extensively investigated lepton mixing patterns that has emerged through the use of flavor symmetries is the tribimaximal (TBM) mixing pattern \cite{Harrison:2002er, Xing:2002sw, Harrison:2002kp}:
\begin{equation}
U_{TBM} = \left(
\begin{array}{ccc}
-\frac{\sqrt{2}}{\sqrt{3}} & \frac{1}{\sqrt{3}} & 0 \\ 
\frac{1}{\sqrt{6}} & \frac{1}{\sqrt{3}} & -\frac{1}{\sqrt{2}}\\ 
\frac{1}{\sqrt{6}} & \frac{1}{\sqrt{3}} & \frac{1}{\sqrt{2}} 
\end{array}
\right).
\end{equation} 
One of the key predictions of the  TBM mixing is that the reactor mixing angle $\theta_{13}$ should be zero, while the atmospheric mixing angle is expected to be maximal, i.e., $\theta_{23} = \frac{\pi}{4}$. Additionally, TBM mixing predicts a specific value for the solar mixing angle, $\theta_{12} = \sin^{-1}(1/\sqrt{3})$. The transformation matrices for the TBM mixing are
\begin{equation}
G_1 = \frac{1}{3} \begin{pmatrix}
    1     &  -2 & 2 \\
   -2    &  -2  &  -1\\
   2      & -1 &  -2
\end{pmatrix},
\end{equation}
\begin{equation}
G_2 = \frac{1}{3} \begin{pmatrix}
    -1     &  2 & -2 \\
   2    &  -1 &  -2\\
   -2      & -2 &  -1
\end{pmatrix},
\end{equation}
and
\begin{equation}
G_3 = \begin{pmatrix}
    -1     &  0 & 0 \\
   0   &  0  &  1\\
   0      & 1 &  0
\end{pmatrix}.
\end{equation}

Even though recent neutrino oscillation experiments have established a non-zero value for the reactor mixing angle $\theta_{13}$ \cite{T2K:2011ypd, MINOS:2011amj, DoubleChooz:2011ymz, DayaBay:2012fng, RENO:2012mkc}, it is still possible to explain the experimental data by considering a mixing matrix that retains only the first or the second column of the TBM mixing matrix. Such mixing patterns are known as TM$_1$ or TM$_2$ mixing and have been extensively studied in the literature \cite{Albright:2008rp, Albright:2010ap, He:2006qd, He:2011gb, Lam:2006wm, Xing:2006ms, Kumar:2010qz, Kumar:2013ela}. Unlike TBM mixing, the neutrino mass matrix giving rise to TM$_1$ or TM$_2$ mixing has only $G_1$ or $G_2$ as its residual flavor symmetry, respectively. The mass matrix for TM$_1$ or  TM$_2$ mixing doesn't have $G_3$ residual symmetry and hence can accommodate a non-zero $\theta_{13}$.
 
The study of residual flavor symmetries in the neutrino mass matrix has inspired the investigation of other types of symmetries as well. One such symmetry is the generalized remnant CP transformation \cite{Ecker:1981wv, Neufeld:1987wa, Grimus:1995zi, Branco:2011zb, Feruglio:2012cw}. Unlike the residual flavor symmetries, these transformations do not leave the neutrino mass matrix $M_{\nu}$ invariant, but rather transform it to its complex conjugate $M_{\nu}^{\ast}$. The generalized CP transformations are denoted by $X_i$ and are related to the lepton mixing matrix $U$ through the relation 
\begin{equation}
X_i = U d_i U^T \text{  for } i=1,2,3.
\end{equation} 
The generalized CP symmetry $X_i$ is present in $M_{\nu}$ if 
\begin{equation}
X^T_i M_{\nu} X_i = M_{\nu}^{\ast}.
\end{equation}
Since, the TBM mixing matrix is real, we note that $G_i=X_i$ mathematically. 
Furthermore the three transformations $G_i$ or $X_i$ along with the identity transformation form the group $Z_2 \times Z_2$. Starting form TBM neutrino mixing matrix, one can construct a neutrino mass matrix that exhibits one of the three independent generalized CP symmetries (X$_1$, X$_2$, or $X_3$). The imposition of any two of these symmetries leads to variants of TBM mixing possessing a single residual flavor symmetry. In other words, performing two CP transformations $X_i X_j^*$ is equivalent to a flavor symmetry transformation $G_k$, where $i$, $j$, and $k$ are cyclic permutations of ${1,2,3}$. This observation is important as it allows to explain the lepton mixing matrix in terms of the remnant CP symmetries. Additionally, imposing generalized CP symmetries instead of a flavor symmetry also puts constraints on the Majorana CP phases, making it an advantageous approach \cite{Chen:2014wxa}. Owing to its significance,  this approach has been systematically studied \cite{Feruglio:2012cw, Chen:2014wxa, Feruglio:2013hia, Ding:2013bpa, Chen:2014tpa, Hagedorn:2014wha, Ding:2014ora, Everett:2015oka, Chen:2015nha, Girardi:2015rwa, Chen:2016ica, King:2017guk, Lu:2018oxc, Barreiros:2018bju, Nath:2018fvw, Chen:2018zbq, Chen:2019fgb, Hagedorn:2021ldq}.

The imposition of a single residual CP-symmetry also allows us to obtain variants of TBM mixing. Just like the variants TM$_1$ and TM$_2$ of TBM mixing, these variants can also exhibit non-zero $\theta_{13}$ and CP violation. Therefore, such variants become relevant for current and forthcoming experimental data. For TBM mixing, the generalized CP symmetry $X_3$ is known as the generalized $\mu-\tau$ symmetry \cite{Xing:2015fdg, Mohapatra:2015gwa, Chen:2015siy} and has received more attention than the other two CP symmetries. Nevertheless, the phenomenology of all three symmetries has been investigated \cite{Chen:2018zbq}. 

Apart from the residual flavour symmetries and remnant CP symmetries, alternative approaches such as imposing texture zeros \cite{Frampton:2002yf, Xing:2002ta, Desai:2002sz, Grimus:2004az, Dev:2006xu, Dev:2006qe, Ahuja:2007vh, Kumar:2011vf, Fritzsch:2011qv, Meloni:2014yea, Dev:2014dla, Singh:2016qcf, Razzaghi:2022jgq}, vanishing cofactors \cite{Lavoura:2004tu, Ma:2005py, Lashin:2007dm, Lashin:2009yd, Dev:2010if, Dev:2011hf, Araki:2012ip, Liao:2013saa, Wang:2014dka, Singh:2022ijf, Dey:2023tsk}, hybrid textures \cite{Kaneko:2005yz, Dev:2009he, Wang:2013woa, Dev:2013nua, Dev:2020xzq, Ankush:2021opd}, or equalities among the neutrino mass matrix elements \cite{Dev:2013xca} have also been explored to explain the neutrino masses and mixings. Assuming neutrinos to be Majorana particles and charged lepton mass matrix to be diagonal, the presence of texture zeros in the neutrino mass matrix is one of the simplest possibilities consistent with the current neutrino data. There are fifteen possible patterns of two texture zeros in the neutrino mass matrix, but only seven of them are compatible with the current neutrino oscillation data (Table \ref{tab2tz}). These seven patterns can be classified into three classes (A, B, and C) depending on the type of the mass hierarchy \cite{Frampton:2002yf}. 

The simultaneous use of both flavor symmetry and texture zeros or vanishing cofactors in constructing neutrino mass matrices has resulted in highly predictive patterns. For example, incorporation of texture zeros or vanishing cofactors along with the mixing patterns TM$_1$ or  TM$_2$ are compatible with the current data and testable with the future data \cite{Gautam:2016qyw, Kumar:2017hjn, Gautam:2018izb, Zhao:2020cjm, Mazumder:2022ywa}.

Another interesting possibility is to impose texture zeros in conjunction with the generalized CP symmetries. For instance, the effects of texture zeros or vanishing cofactors in conjunction with the generalized CP symmetry $X_3$ corresponding to TBM mixing have been studied in literature \cite{Nishi:2016wki, Nishi:2018vlz}.

In this work, we aim to investigate the consequences of incorporating texture zeros into the framework of the generalized CP symmetries X$_1$ and X$_2$ corresponding to TBM mixing. By combining these two approaches together, we expect to obtain even more predictive neutrino mass matrices. Specifically, we will explore the possible patterns of texture zeros in the neutrino mass matrix that are consistent with the symmetries imposed by X$_1$ and X$_2$. Our analysis will also consider the implications of these patterns for the values of the neutrino masses, mixing angles, and the CP-violating phases. The form of neutrino mixing matrix and mass matrix for X$_1$ and X$_2$ CP symmetries is derived in Section \ref{sec2} where we also show the way to obtain the predictions for masses, mixing angles, and CP-violating phases. We study the implications of imposing the one and two texture zero conditions with X$_1$ and X$_2$ CP symmetries and study the implications for neutrino parameters in light of the current experimental data. We summarize the results of the study in Section \ref{sec3} and conclusions in Section \ref{sec4}.

\section{Formalism and Methodology \label{sec2}}

\subsection{Generalized CP Symmetry \label{secgcp}}

Neutrino mass matrices invariant under X$_1$ or X$_2$ generalized CP symmetries can be written as
\begin{equation}\label{mx1}
M_{X_1} = \left(
\begin{array}{ccc}
 a+4 i (x+y) & 2 b+2 i x & 2 c+2 i y \\
 2 b+2 i x & 4 b+d+i (-3 x-y) & a-b-c-d+2 i (-x-y) \\
 2 c+2 i y & a-b-c-d+2 i (-x-y) & 4 c+d+i (-x-3 y)
\end{array}
\right),
\end{equation}
\begin{equation}\label{mx2}
M_{X_2} = \left(
\begin{array}{ccc}
 a+2 i (x+y) & b+i x & c+i y \\
 b+i x & d-2 i y & a+c-d+i (-x-y) \\
 c+i y & a+c-d+i (-x-y) & b-c+d-2 i x
\end{array}
\right),
\end{equation}
where, $a, b, c, d, x$ and $y$ are real parameters. The mass matrix with X$_1$ CP symmetry ($M_{X_1}$) is diagonalized by the lepton mixing matrix \cite{Chen:2018zbq}
\begin{equation}
U_{X_1} = U_{TBM} \text{diag} (1,i,i) O_{\nu} Q_{\nu}
\end{equation}
and the mass matrix with X$_2$ CP symmetry ($M_{X_2}$) is diagonalized by the lepton mixing matrix
\begin{equation}
U_{X_2} = U_{TBM} \text{diag} (i,1,i) O_{\nu} Q_{\nu}.
\end{equation}
Here, the matrix $O_{\nu}$ contains the three angles ($\theta_1$, $\theta_2$, $\theta_3$) that parametrize the mixing matrix $U$ for a particular generalized CP symmetry and is given as
\begin{small}
\begin{equation} \label{eq:onu}
O_{\nu} = \begin{pmatrix}
    1       & 0 & 0 \\
    0       & \cos \theta_1 & \sin \theta_1 \\
   0      & -\sin \theta_1 &  \cos \theta_1
\end{pmatrix}
\begin{pmatrix}
    \cos \theta_2      & 0 & \sin \theta_2 \\
    0       & 1 & 0 \\
   -\sin \theta_2      & 0 &  \cos \theta_2
\end{pmatrix}
\begin{pmatrix}
    \cos \theta_3       &  \sin \theta_3 & 0 \\
     -\sin \theta_3     &  \cos \theta_3 & \\
   0      & 0 &  1
\end{pmatrix}.
\end{equation}
\end{small}
The diagonal matrix of CP parities $Q_{\nu}$ is given by
\begin{equation} \label{eq:qnu}
Q_{\nu} = \textrm{diag} (e^{ik_1 \pi/2},e^{ik_2 \pi/2},e^{ik_3 \pi/2}),
\end{equation}
with integer values of $k_i$ ($i=1,2,3$).
The mixing angles ($\theta_{12}$, $\theta_{23}$, and $\theta_{13}$) can be calculated from absolute values of the respective entries of $U$ and the CP-violating phases can be calculated from the three CP invariants \cite{Branco:1986gr, Jenkins:2007ip, Branco:2011zb}  by comparing the values obtained here with their counterparts in the PDG parametrization \cite{ParticleDataGroup:2018ovx} [Cf. Eq. (\ref{eqpdg})]. The Jarlskog invariant $J_{CP}$ is given as \cite{Jarlskog:1985ht}
\begin{equation}
J_{CP}  =\text{Im}  \left[ U_{11}U_{33} U^*_{13} U^*_{31}  \right].
\end{equation}
The invariants $I_1$ and $I_2$ associated with the Majorana phases are given as \cite{Branco:1986gr, Jenkins:2007ip, Branco:2011zb}
\[
I_1 = \text{Im} \left[ U_{12}^2 U_{11}^{*2} \right] 
\]
and
\[
I_2 = \text{Im} \left[ U_{13}^2 U_{11}^{*2} \right] .
\]

\subsection{Two Texture Zeros}

Assuming neutrinos to be Majorana particles and taking the charged lepton mass matrix as diagonal, there are fifteen potential patterns of two texture zeros in the neutrino mass matrix. However, only seven of them are compatible with current neutrino oscillation data. These seven patterns have been summarized in Table \ref{tab2tz}.

\begin{table}[tb]
\begin{center}
\begin{tabular}{cc}
\hline
 Type  &  Constraining Equations    \\
 \hline
 A$_1$ &  $M_{ee}=0$, $M_{e\mu}=0$     \\
 A$_2$ &  $M_{ee}=0$, $M_{e\tau}=0$     \\
 B$_1$ &  $M_{e\tau}=0$, $M_{\mu\mu}=0$   \\
 B$_2$ &  $M_{e\mu}=0$, $M_{\tau\tau}=0$  \\
 B$_3$ &  $M_{e\mu}=0$, $M_{\mu\mu}=0$   \\
 B$_4$ &  $M_{e\tau}=0$, $M_{\tau\tau}=0$  \\
 C   & $M_{\mu\mu}=0$, $M_{\tau\tau}=0$  \\
 \hline
\end{tabular}
\end{center}
\caption{Viable two texture zero neutrino mass matrices classified 
into three classes.}
\label{tab2tz}
\end{table}

\subsection{Two Texture Zeros with X$_1$ or X$_2$ Generalized CP Symmetry \label{sec2.3}}

Imposing the X$_1$ generalized CP symmetry on the seven neutrino mass matrices with two texture zeros, we obtain
\begin{equation}\label{a1x1}
M^{A_1}_{X_1} = \left(
\begin{array}{ccc}
 0 & 0 & 2 c \\
 0 & d & -c-d \\
 2 c & -c-d & 4 c+d
\end{array}
\right),
\end{equation}
\begin{equation}\label{a2x1}
M^{A_2}_{X_1}  = \left(
\begin{array}{ccc}
 0 & 2 b & 0 \\
 2 b & 4 b+d & -b-d \\
 0 & -b-d & d
\end{array}
\right),
\end{equation}
\begin{equation}\label{b1x1}
M^{B_1}_{X_1}  = 
\left(
\begin{array}{ccc}
 a & 2 b & 0 \\
 2 b & 0 & a+3 b \\
 0 & a+3 b & -4 b
\end{array}
\right),
\end{equation}
\begin{equation}\label{b2x1}
M^{B_2}_{X_1}  = 
\left(
\begin{array}{ccc}
 a & 0 & 2 c \\
 0 & -4 c & a+3 c \\
 2 c & a+3 c & 0
\end{array}
\right),
\end{equation}
\begin{equation}\label{b3x1}
M^{B_3}_{X_1} = 
\left(
\begin{array}{ccc}
 a & 0 & 2 c \\
 0 & 0 & a-c \\
 2 c & a-c & 4 c
\end{array}
\right),
\end{equation}
\begin{equation}\label{b4x1}
M^{B_4}_{X_1}  = 
\left(
\begin{array}{ccc}
 a & 2 b & 0 \\
 2 b & 4 b & a-b \\
 0 & a-b & 0
\end{array}
\right), \text{ and}
\end{equation}
\begin{equation}\label{cx1}
M^{C}_{X_1} = 
\left(
\begin{array}{ccc}
 a & 2 b & 2 b \\
 2 b & 0 & a+2 b \\
 2 b & a+2 b & 0
\end{array}
\right)
\end{equation}
where, $a, b, c, d$ are real parameters. If we impose the X$_2$ CP symmetry, the seven possibilities for the neutrino mass matrix are
\begin{equation}\label{a1x2}
M^{A_1}_{X_2}  = \left(
\begin{array}{ccc}
 0 & 0 & c \\
 0 & c+d & -d \\
 c & -d & d
\end{array}
\right)
\end{equation}
\begin{equation}\label{a2x2}
M^{A_2}_{X_2}  = \left(
\begin{array}{ccc}
 0 & b & 0 \\
 b & d & -d \\
 0 & -d & b+d
\end{array}
\right),
\end{equation}
\begin{equation}\label{b1x2}
M^{B_1}_{X_2}  = \left(
\begin{array}{ccc}
 a+2 i x & b+i x & 0 \\
 b+i x & 0 & a-i x \\
 0 & a-i x & b-2 i x
\end{array}
\right),
\end{equation}
\begin{equation}\label{b2x2}
M^{B_2}_{X_2} = \left(
\begin{array}{ccc}
 a+2 i y & 0 & c+i y \\
 0 & c-2 i y & a-i y \\
 c+i y & a-i y & 0
\end{array}
\right),
\end{equation}
\begin{equation}\label{b3x2}
M^{B_3}_{X_2} = \left(
\begin{array}{ccc}
 a & 0 & c \\
 0 & 0 & a+c \\
 c & a+c & -c
\end{array}
\right),
\end{equation}
\begin{equation}\label{b4x2}
M^{B_4}_{X_2} = \left(
\begin{array}{ccc}
 a & b & 0 \\
 b & -b & a+b \\
 0 & a+b & 0
\end{array}
\right), \text{ and}
\end{equation}
\begin{equation}\label{cx2}
M^{C}_{X_2} = \left(
\begin{array}{ccc}
 a & b & b \\
 b & 0 & a+b \\
 b & a+b & 0
\end{array}
\right)
\end{equation}
where, $a, b, c, d, x$ and $y$ are real parameters. 

We observe that for the neutrino mass matrix with X$_1$ CP symmetry and two texture zeros, all the seven classes become real. However,  for the neutrino mass matrices with X$_2$ CP symmetry and two texture zeros, the classes B$_1$ and $ B_2$ are complex while other 5 classes are real. We know that the real parts of the mass matrices remain invariant under X$_1$ or X$_2$ CP symmetry and are also invariant under $G_1$ or $G_2$ flavor symmetry. Hence, they can be diagonalized by TM$_1$ or TM$_2$ mixing matrices, respectively. Thus, all these mass matrices, except those for classes $B_1, B_2$ with X$_2$ CP symmetry, can be diagonalized either by the TM$_1$ or TM$_2$ mixing matrices. In other words, all the seven cases of two texture zero  mass matrices with X$_1$ CP symmetry are diagonalized by the TM$_1$ mixing matrix
\begin{equation}\label{utm1}
U_{TM_1} = \left(
\begin{array}{ccc}
  \frac{2}{\sqrt{3}} & \frac{1}{\sqrt{3}} \cos \theta &
\frac{1}{\sqrt{3}} \sin \theta \\
  -\frac{1}{\sqrt{6}} &
      \frac{1}{\sqrt{3}}\cos\theta-\frac{
                        \sin \theta}{\sqrt{2}} &
 \frac{1}{\sqrt{3}}\sin \theta
                                                 +\frac{\cos \theta}{\sqrt{2}} \\
    -\frac{1}{\sqrt{6}} &
      \frac{1}{\sqrt{3}}\cos\theta+\frac{ 
                        \sin \theta}{\sqrt{2}} &
 \frac{1}{\sqrt{3}}\sin \theta
-\frac{ \cos \theta}{\sqrt{2}}
\end{array}
\right)P.
\end{equation}
and the five cases of the two texture zero mass matrices with X$_2$ CP symmetry (classes A$_1$, A$_2$, B$_3$, B$_4$, and C) are diagonalized by the TM$_2$ mixing matrix
\begin{equation}\label{utm2}
U_{TM_2} = \left(
\begin{array}{ccc}
 -\sqrt{\frac{2}{3}} \cos\theta & \frac{1}{\sqrt{3}} & -\sqrt{\frac{2}{3}} \sin
  \theta \\
 \frac{\cos\theta}{\sqrt{6}}+\frac{\sin\theta}{\sqrt{2}} &
   \frac{1}{\sqrt{3}} & \frac{\sin\theta}{\sqrt{6}}-\frac{\cos (\theta
   )}{\sqrt{2}} \\
 \frac{\cos\theta}{\sqrt{6}}-\frac{\sin\theta}{\sqrt{2}} &
   \frac{1}{\sqrt{3}} & \frac{\cos\theta}{\sqrt{2}}+\frac{\sin (\theta
   )}{\sqrt{6}}
\end{array}
\right)P.
\end{equation}
The mass matrices for classes B$_1$ and B$_2$ with X$_2$ CP symmetry need to be diagonalized with  another method that we will describe later in this section. 

The condition of two texture zeros in the neutrino mass matrix at $(p,q)$ and $(r,s)$ positions, when imposed with X$_1$ CP symmetry, implies:
\begin{equation}
(M_{X_1})_{pq} = 0 \ \textrm{and} \  (M_{X_1})_{rs} = 0. 
\end{equation}
When expressed in terms of the mass eigenvalues and the mixing matrix elements, these two complex equations become
\begin{equation}
m_1 U_{p1}U_{q1}e^{2i\phi_1} + m_2 U_{p2}U_{q2}e^{2i\phi_2} + m_3 U_{p3}U_{q3}e^{2i\phi_3} = 0 \label{eqzero1}
\end{equation}
and
\begin{equation}
m_1 U_{r1}U_{s1}e^{2i\phi_1} + m_2 U_{r2}U_{s2}e^{2i\phi_2} + m_3 U_{r3}U_{s3}e^{2i\phi_3} = 0 \label{eqzero2}
\end{equation}
where $p, q, r$ and $s$ can take the values $1$, $2$, and $3$. These equations can be simultaneously solved to obtain two complex mass ratios:
\begin{equation}
\frac{m_1}{m_2}e^{2i(\phi_1 - \phi_2)}=\frac{U_{r2}U_{s2}U_{p3}U_{q3}-U_{p2}U_{q2}U_{r3}U_{s3}}{U_{p1}U_{q1}U_{r3}U_{s3}-U_{p3}U_{q3}U_{r1}U_{s1}} \label{eqratio1}
\end{equation}
and
\begin{equation}
\frac{m_1}{m_3}e^{2i(\phi_1 - \phi_3)}=\frac{U_{r3}U_{s3}U_{p2}U_{q2}-U_{p3}U_{q3}U_{r2}U_{s2}}{U_{p1}U_{q1}U_{r2}U_{s2}-U_{p2}U_{q2}U_{r1}U_{s1}} . \label{eqratio2}
\end{equation}
The magnitudes of the two mass ratios in Eqs. (\ref{eqratio1}) and (\ref{eqratio2}), are defined as
\begin{equation}
\textrm{r}_{12}=\left|\frac{m_1}{m_2}e^{2i(\phi_1 - \phi_2)}\right|, \ \ \ \ \ \ 
\textrm{r}_{13}=\left|\frac{m_1}{m_3}e^{2i(\phi_1 - \phi_3)}\right| .\label{eqratiomod}
\end{equation}

The magnitudes of the two mass ratios can be used to obtain the expression for the parameter $R_\nu$ defined as the ratio of mass squared differences:
\begin{equation}\label{eq:rnu1}
R_\nu \equiv \frac{\Delta m_{21}^2}{\left|\Delta m_{31}^2\right|} = \frac{\left(\frac{1}{\text{r}_{12}}\right)^2-1}{\left|\left(\frac{1}{\text{r}_{13}}\right)^2-1\right|}
\end{equation} where $m_1 < m_3$ for NH ($m_1 > m_3$ for IH). The two CP-violating Majorana phases $\phi_{12} \equiv (\phi_1 - \phi_2)$ and $\phi_{13} \equiv (\phi_1 - \phi_3)$ are given by
\begin{align}
\phi_{12}  & =-\frac{1}{2}\textrm{Arg}\left(\frac{U_{r2}U_{s2}U_{p3}U_{q3}-U_{p2}U_{q2}U_{r3}U_{s3}}{U_{p1}U_{q1}U_{r3}U_{s3}-U_{p3}U_{q3}U_{r1}U_{s1}}\right), \\
\phi_{13}  & =-\frac{1}{2}\textrm{Arg}\left(\frac{U_{r3}U_{s3}U_{p2}U_{q2}-U_{p3}U_{q3}U_{r2}U_{s2}}{U_{p1}U_{q1}U_{r2}U_{s2}-U_{p2}U_{q2}U_{r1}U_{s1}}\right).
\end{align}
Since the mass matrices in the present case are real, the two Majorana type CP phases take trivial values of $0$, $\frac{\pi}{2}$ or $\pi$. For the neutrino mass matrix with two texture zeros and X$_1$ CP symmetry to be compatible with the present neutrino oscillation data, the values of $R_\nu$ obtained from Eq. (\ref{eq:rnu1}) should lie within its experimentally allowed range. The mixing matrix used in the above equations is $U_{TM_1}$. If we replace it with $U_{TM_2}$, the same procedure can be used to study the neutrino mass matrix with two texture zeros and X$_2$ CP symmetry.

The neutrino mass matrices corresponding to classes $B_1, B_2$ with X$_2$ CP symmetry are complex and follow the symmetry relations
\begin{equation}
X_{2}^T M^{B_1} X_2 =  (M^{B_1})^*
\end{equation}
and
\begin{equation}
X_{2}^T M^{B_2} X_2 = (M^{B_2})^*.
\end{equation}
To diagonalize these mass matrices, we first make them real using the transformation
\begin{equation}
M^{B_j}_{\textrm{real}} =  \textrm{diag}(i,1,i)^T U_{TBM}^T M^{B_j} U_{TBM} \textrm{diag}(i,1,i)
\end{equation}
where $j=1$ or $2$. The resulting real symmetric mass matrix is diagonalizable by an orthogonal matrix $O$ as
\begin{equation}
\left(M^{B_j}\right)_{\textrm{diag}} =  O^T M^{B_j}_{\textrm{real}} O
\end{equation}
that can be parametrized in terms of three neutrino masses ($m_1$, $m_2$, $m_3$) using the  three invariants 
\begin{equation}
Tr [M_{\textrm{real}}] = m_1 + m_2 + m_3,
\end{equation},  
\begin{equation}
Tr [(M_{\textrm{real}})^2] = m_1^2 + m_2^2 + m_3^2,
\end{equation}
and
\begin{equation}
Det [M_{\textrm{real}}] = m_1 m_2 m_3.
\end{equation}
By solving above three equations, we can obtain the desired orthogonal mixing matrix $O$ that diagonalizes the neutrino mass matrix of class B$_1$ or B$_2$ with X$_2$ CP symmetry:
\begin{small}
\begin{equation}
O = 
\left(
\begin{array}{ccc}
 \frac{\sqrt{3} m_2
   m_3-C_1}{\left(-m_1^2+m_2^2+m_3^2-m_2 m_3\right)
   N_1} & \frac{\sqrt{3} m_1 m_3-C_1}{\left(m_1^2-m_3
   m_1-m_2^2+m_3^2\right) N_2} & \frac{\sqrt{3} m_1
   m_2-C_1}{\left(m_1^2-m_2 m_1+m_2^2-m_3^2\right)
   N_3} \\
 \frac{\sqrt{2} C_2 \left(-m_1^2+(m_2+m_3)^2+\sqrt{3}
   C_1\right)}{(m_2+m_3)
   \left(-m_1^2+m_2^2+m_3^2-\sqrt{3} C_1-4 m_2 m_3\right)
   N_1} & \frac{\sqrt{2} C_2 \left(\sqrt{3}
   C_1+(m_1-m_2+m_3)
   (m_1+m_2+m_3)\right)}{(m_1+m_3) \left(m_1^2-4
   m_3 m_1-m_2^2+m_3^2-\sqrt{3} C_1\right) N_2} &
   \frac{\sqrt{2} C_2 \left(\sqrt{3} C_1+(m_1+m_2-m_3)
   (m_1+m_2+m_3)\right)}{(m_1+m_2) \left(m_1^2-4
   m_2 m_1+m_2^2-m_3^2-\sqrt{3} C_1\right) N_3} \\
 \frac{1}{N_1} & \frac{1}{N_2} & \frac{1}{N_3}
\end{array}
\right), \label{o}
\end{equation}
\end{small}
where 
\begin{equation}
C_1 = \sqrt{(m_1+m_2-m_3) (m_1-m_2+m_3)
   (-m_1+m_2+m_3) (m_1+m_2+m_3)},
\end{equation}
and
 \begin{equation}
C_2 = \sqrt{-\frac{(m_1+m_2) (m_1+m_3)
   (m_2+m_3)}{m_1+m_2+m_3}}.
\end{equation}
Here, the normalization factors $N_1, N_2$ and $N_3$ are given by
 \begin{equation}
N_1 =  2 \sqrt{\frac{(m_1-m_2) (m_3-m_2)
   (m_1+m_2+m_3)}{(m_2+m_3) \left(-\sqrt{3}
   C_1-m_1^2+m_2^2-4 m_2 m_3+m_3^2\right)}} ,
\end{equation}
\begin{equation}
N_2 = 2 \sqrt{\frac{(m_1-m_2) (m_2-m_3)
   (m_1+m_2+m_3)}{(m_1+m_3) \left(-\sqrt{3}
   C_1+m_1^2-4 m_1 m_3-m_2^2+m_3^2\right)}} ,
\end{equation}
and
\begin{equation}   
N_3 = 2 \sqrt{\frac{(m_1-m_3) (m_3-m_2)
   (m_1+m_2+m_3)}{(m_1+m_2) \left(-\sqrt{3}
   C_1+m_1^2-4 m_1 m_2+m_2^2-m_3^2\right)}}.
\end{equation}
Using the above expression of the orthogonal matrix $O$ (Eq. \ref{o}), the full mixing matrix that diagonalizes the neutrino mass matrices $M^{B_1}_{X_2}$ and $M^{B_2}_{X_2}$ can be written as
\begin{equation}
U = U_{TBM} \text{diag} (i,1,i) O Q_{\nu}.
\label{diagonalizingu}
\end{equation}
This matrix $U$ depends upon the neutrinos masses $m_1$, $m_2$, and $m_3$ and can be used to calculate the predictions of the patterns $M^{B_1}_{X_2}$ and $M^{B_2}_{X_2}$ for the mixing angles and CP-violating phases.

\subsection{One Texture Zero with X$_1$ or X$_2$ Generalized CP Symmetry \label{sec:2.4}}

The neutrino mass matrices having one texture zero and X$_1$ CP symmetry can be obtained from $M_{X_1}$ in Eq. (\ref{mx1}) by substituting the texture zero condition at the appropriate place:
\begin{equation}
M_{X_1}^{I} = \left(
\begin{array}{ccc}
 0 & 2 b+2 i x & 2 c-2 i x \\
 2 b+2 i x & 4 b+d-2 i x & -b-c-d \\
 2 c-2 i x & -b-c-d & 4 c+d+2 i x
\end{array}
\right),
\end{equation}
\begin{equation}
M_{X_1}^{II} = \left(
\begin{array}{ccc}
 a+4 i y & 0 & 2 c+2 i y \\
 0 & d-i y & a-c-d-2 i y \\
 2 c+2 i y & a-c-d-2 i y & 4 c+d-3 i y
\end{array}
\right),
\end{equation}
\begin{equation}
M_{X_1}^{III} = \left(
\begin{array}{ccc}
 a+4 i x & 2 b+2 i x & 0 \\
 2 b+2 i x & 4 b+d-3 i x & a-b-d-2 i x \\
 0 & a-b-d-2 i x & d-i x
\end{array}
\right),
\end{equation}
\begin{equation}
M_{X_1}^{IV} = \left(
\begin{array}{ccc}
 a-8 i x & 2 b+2 i x & 2 c-6 i x \\
 2 b+2 i x & 0 & a+3 b-c+4 i x \\
 2 c-6 i x & a+3 b-c+4 i x & -4 b+4 c+8 i x
\end{array}
\right),
\end{equation}
and
\begin{equation}
M_{X_1}^{V} = \left(
\begin{array}{ccc}
 a & 2 b+2 i x & 2 c-2 i x \\
 2 b+2 i x & a+3 b-c-2 i x & 0 \\
 2 c-2 i x & 0 & a-b+3 c+2 i x
\end{array}
\right).
\end{equation}

\begin{equation}
M_{X_1}^{VI} = \left(
\begin{array}{ccc}
 a+\frac{8 i x}{3} & 2 b+2 i x & 2 c-\frac{2 i x}{3} \\
 2 b+2 i x & 4 b-4 c-\frac{8 i x}{3} & a-b+3 c-\frac{4 i x}{3} \\
 2 c-\frac{2 i x}{3} & a-b+3 c-\frac{4 i x}{3} & 0
\end{array}
\right),
\end{equation}
where $a$, $b$, $c$, $d$, $x$, and $y$ are real parameters. Using the similar methodology, the neutrino mass matrices having one texture zero and X$_2$ CP symmetry can be obtained from $M_{X_2}$ in Eq. (\ref{mx2}) by substituting the texture zero condition at the appropriate place:
\begin{equation}
M_{X_2}^{I} = \left(
\begin{array}{ccc}
 0 & b+i x & c-i x \\
 b+i x & d+2 i x & c-d \\
 c-i x & c-d & b-c+d-2 i x
\end{array}
\right),
\end{equation}
\begin{equation}
M_{X_2}^{II} = \left(
\begin{array}{ccc}
 a+2 i y & 0 & c+i y \\
 0 & d-2 i y & a+c-d-i y \\
 c+i y & a+c-d-i y & d-c
\end{array}
\right),
\end{equation}
\begin{equation}
M_{X_2}^{III} = \left(
\begin{array}{ccc}
 a+2 i x & b+i x & 0 \\
 b+i x & d & a-d-i x \\
 0 & a-d-i x & b+d-2 i x
\end{array}
\right),
\end{equation}
\begin{equation}
M_{X_2}^{IV} = \left(
\begin{array}{ccc}
 a+2 i x & b+i x & c \\
 b+i x & 0 & a+c-i x \\
 c & a+c-i x & b-c-2 i x
\end{array}
\right),
\end{equation}
\begin{equation}
M_{X_2}^{V} = \left(
\begin{array}{ccc}
 a & b+i x & c-i x \\
 b+i x & a+c+2 i x & 0 \\
 c-i x & 0 & a+b-2 i x
\end{array}
\right),
\end{equation}
and
\begin{equation}
M_{X_2}^{VI} = \left(
\begin{array}{ccc}
 a+2 i y & b & c+i y \\
 b & -b+c-2 i y & a+b-i y \\
 c+i y & a+b-i y & 0
\end{array}
\right).
\end{equation}

The condition of one texture zero in the neutrino mass matrix when imposed along with X$_1$ CP symmetry, implies: 
\begin{equation}
(M_{X_1})_{jk} = 0.
\end{equation}
The above condition yields a complex equation given by
\begin{equation}
m_1 P + m_2 Q + m_3 R = 0
\end{equation}
with the tree complex coefficients  $P = U_{j1}U_{k1} e^{2 i \phi_1}$, $Q = U_{j2}U_{k2} e^{2 i \phi_2}$, and $R = U_{j3}U_{k3} e^{2 i \phi_3}$. Here, the indices $j$ and $k$ can take values $1$, $2$ and $3$. This complex equation can be used to calculate two mass ratios $\frac{m_1}{m_2}$ and $\frac{m_1}{m_3}$ :
\begin{equation}\label{eqm12}
\frac{m_1}{m_2} = \frac{\textrm{Re}(R) \textrm{Im}(Q)-\textrm{Re}(Q) \textrm{Im}(R)}{\textrm{Re}(P) \textrm{Im}(R)-\textrm{Re}(R) \textrm{Im}(P)}
\end{equation}
and
\begin{equation}\label{eqm13}
\frac{m_1}{m_3} = \frac{\textrm{Re}(R) \textrm{Im}(Q)-\textrm{Re}(Q) \textrm{Im}(R)}{\textrm{Re}(Q) \textrm{Im}(P)-\textrm{Re}(P) \textrm{Im}(Q)}
\end{equation}
where the symbols Re and Im denote the real and imaginary parts, respectively.

We use the above mass ratios to obtain the expression for the parameter $R_\nu$ defined as 
\begin{equation}\label{eq:rnu}
R_\nu \equiv \frac{\Delta m_{21}^2}{|\Delta m_{31}^2|} = \frac{(\frac{m_2}{m_1})^2-1}{|(\frac{m_3}{m_1})^2-1|}.
\end{equation} 
For a neutrino mass matrix with one texture zero and X$_1$ CP symmetry to be compatible with the present neutrino oscillation data, the values of $R_\nu$ obtained from Eq. (\ref{eq:rnu}) should lie within its experimentally allowed range. Similarly, one can obtain mass ratios and $R_{\nu}$ for 
neutrino mass matrices with one texture zero and X$_2$ CP symmetry and analyze their predictions.

\begin{table*}[t]
\begin{center}
\begin{tabular}{|c|c|c|}
 \hline
Parameter & Normal Ordering & Inverted Ordering \\
  & best fit $\pm 1 \sigma$ ~~ $3 \sigma$ range  & best fit $\pm 1 \sigma$~~~~~~ $3 \sigma$ range \\
 \hline 
$\theta_{12}^{\circ}$ & $34.3^{+1.0}_{-1.0}$ ~ $31.4$ - $37.4$ & $34.3^{+1.0}_{-1.0}$ ~~~ $31.4$ - $37.4$ \\
$\theta_{23}^{\circ}$ & $49.26^{+0.79}_{-0.79}$  ~~$41.20$ - $51.33$ & $49.46^{+0.60}_{-0.97}$ ~~~~~~ $41.16$ - $51.25$ \\
$\theta_{13}^{\circ}$ & $8.53^{+0.13}_{-0.12}$  ~~$8.13$ - $8.92$ & $8.58^{+0.12}_{-0.14}$ ~~~~~~ $8.17$ - $8.96$ \\
$\delta_{CP}^{\circ}$ & $194^{+24}_{-22}$ ~~ $128$ - $359$ &  $284^{+26}_{-28}$ ~~~~~~ $200$ - $353$ \\
$\Delta m^{2}_{21}/10^{-5} eV^2 $ & $7.50^{+0.22}_{-0.20}$ ~~$6.94$ - $8.14$ & $7.50^{+0.22}_{-0.20}$ ~~$6.94$ - $8.14$ \\
$|\Delta m^{2}_{31}|/10^{-3} eV^2 $ & $2.55^{+0.02}_{-0.03}$ ~~ $2.47$ - $2.63$ & $2.45^{+0.02}_{-0.03}$ ~~~~~~ $2.37$ - $2.53$ \\
 \hline 
 \end{tabular}
\caption{Current neutrino oscillation parameters from global fits \cite{deSalas:2020pgw}.}
\label{tabdat}
\end{center}
\end{table*}

\section{Analysis, Results, and Discussion \label{sec3}}
We have 7 patterns of two texture zeros with X$_1$ generalized CP symmetry, 7 patterns of two texture zeros with X$_2$ CP symmetry, 6 patterns of one texture zeros with X$_1$ CP symmetry, and 6 patterns of one texture zeros with X$_2$ CP symmetry. Each of these patterns can be studied for NH as well as IH. Therefore, we have 14 patterns for two texture zeros with one generalized CP symmetry and 12 patterns for one texture zero with one generalized CP symmetry. Therefore, we have 26 patterns for NH and 26 patterns for IH.

\subsection{Two texture zeros with one generalized CP symmetry}
We study the neutrino mass matrices with two texture zeros and one generalized CP symmetry using the procedure derived in the previous section. When we calculate the predictions of these patterns for the mixing angles and mass squared differences, we find that they all are disallowed except for patterns $M^{A_1}_{X_2}$ and $M^{A_2}_{X_2}$. In summary, all the seven cases of two texture zeros with X$_1$ CP symmetry [Eqs. (\ref{a1x1})-(\ref{cx1})] are disallowed by the current data for both the hierarchies. Out of the seven patterns of neutrino mass matrix with X$_2$ CP symmetry [Eqs. (\ref{a1x2})-(\ref{cx2})], only patterns A$_1$ and A$_2$ with NH are allowed.  

We first discuss the inconsistency of two texture zeros with $X_1$ CP symmetry. 
We find that all patterns of two texture zeros with $X_1$ CP symmetry are inconsistent with the present experimental data at 3$\sigma$ Confidence Level (CL). The expression of $R_{\nu}$ for pattern $A_1$ with $X_1$ CP symmetry is given by
\begin{equation}
{R_{\nu}}^{A_1}_{X_1} = \frac{\left(\sqrt{6} \tan (\theta )-2\right)^2-1}{\left(-\sqrt{6} \cot (\theta)-2\right)^2-1}.
\end{equation}
The reactor mixing angle $\theta_{13}$ always remains below 6$^\circ$ for the values of parameter $\theta$ (Eq. \ref{utm1}) corresponding to the experimentally allowed 3$\sigma$ range of $R_{\nu}$. Hence, this pattern is inconsistent with the present experimental data at 3$\sigma$ CL. Similarly, the expression of $R_{\nu}$ for pattern $B_1$ with $X_1$ CP symmetry is given by
\begin{equation}
{R_{\nu}}^{B_1}_{X_1} = -\frac{4 \sin (2 \theta )-3 \sqrt{6} \cos (2 \theta )+\sqrt{6}}{-4 \sin (2 \theta )+3 \sqrt{6} \cos (2 \theta )+\sqrt{6}}.
\end{equation}
From this $R_{\nu}$, we find that  $\theta_{13}$ always remains above 12$^\circ$ for the physical values of $\theta$ (that give $R_\nu$ in the experimentally allowed 3$\sigma$ range). Hence, this pattern is also inconsistent with the present experimental data at 3$\sigma$ CL.  The expression of $R_{\nu}$ for pattern $B_3$ with $X_1$ CP symmetry is given by
\begin{equation}
{R_{\nu}}^{B_3}_{X_1} = \frac{-12 \sin (2 \theta )+\sqrt{6} \cos (2 \theta )+5 \sqrt{6}}{-12 \sin (2 \theta
   )+\sqrt{6} \cos (2 \theta )-5 \sqrt{6}}.
\end{equation} 
This expression gives the value of $R_{\nu}$ consistent with its experimentally allowed range at 3$\sigma$ only for $\theta_{13}$ greater than 20$^{\circ}$, which excludes this pattern as well.

Patterns $A_2$, $B_2$ and $B_4$ with $X_1$ CP symmetry are also disallowed because their predictions are related to patterns $A_1$, $B_1$ and $B_3$, respectively, through 2-3 permutation symmetry \cite{Fritzsch:2011qv} which results in same predictions of $R_{\nu}$ and $\theta_{13}$ for these patterns. 
When we impose the $X_1$ CP symmetry on pattern C of two texture zeros, the resulting neutrino mass matrix becomes $\mu-\tau$ symmetric or invariant under the flavor symmetry $G_3$. So, the pattern $M^{C}_{X_1}$ predicts vanishing $\theta_{13}$ and is disallowed. 

All the patterns except $A_1$ and $A_2$ of two texture zeros with $X_2$ CP symmetry are inconsistent with the present experimental data at 3$\sigma$ CL. The expression of $R_{\nu}$ for pattern $B_3$ with $X_2$ CP symmetry is given by
\begin{equation}
{R_{\nu}}^{B_3}_{X_2} = \frac{1}{4} \left(-\sqrt{3} \sin (2 \theta )+\cos (2 \theta )+2\right).
\end{equation}
For the values of $\theta$ (Eq. \ref{utm2}) corresponding to the 3$\sigma$ experimentally allowed ranges of $R_{\nu}$, the reactor mixing angle $\theta_{13}$ always remains above 35$^\circ$. Hence, this pattern is inconsistent with the present experimental data at 3$\sigma$ CL. Pattern $B_4$ with $X_2$ CP symmetry is related to pattern $B_3$ through 2-3 permutation symmetry, and hence it is also disallowed. Pattern C of two texture zeros with $X_2$ CP symmetry results in $\mu-\tau$ symmetric neutrino mass matrix which predicts vanishing $\theta_{13}$ and is disallowed. Patterns $B_1$ and $B_2$ of two texture zeros with $X_2$ CP symmetry have complex mass matrices as shown in Eqs. \ref{b1x2} and \ref{b2x2}. The diagonalization procedure for these patters is given in Section \ref{sec2.3}. Following that procedure, we find that these two patterns are also inconsistent with the present neutrino oscillation data at 3$\sigma$ CL.

\begin{figure}
\centering
\begin{subfigure}{0.4\textwidth}
    \includegraphics[width=\textwidth]{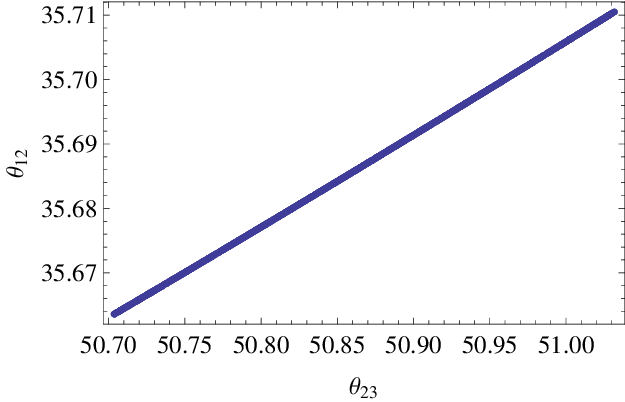}
    \caption{$\theta_{12}$ and $\theta_{23}$ correlation for the class A$_1$ with X$_2$ CP symmetry}   
    \end{subfigure}
\begin{subfigure}{0.4\textwidth}
    \includegraphics[width=\textwidth]{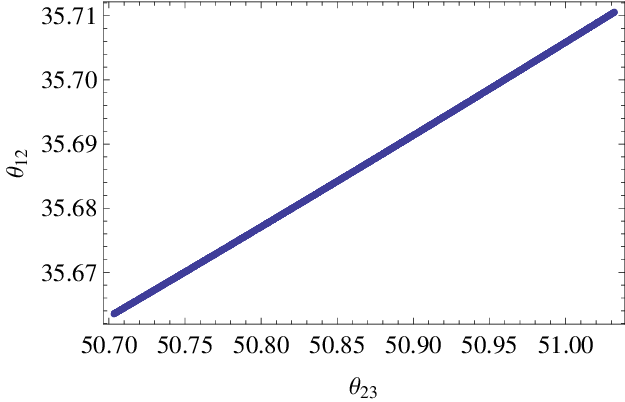}
    \caption{$\theta_{12}$ and $\theta_{23}$ correlation for the class A$_2$ with X$_2$ CP symmetry}     
\end{subfigure}
\begin{subfigure}{0.4\textwidth}
    \includegraphics[width=\textwidth]{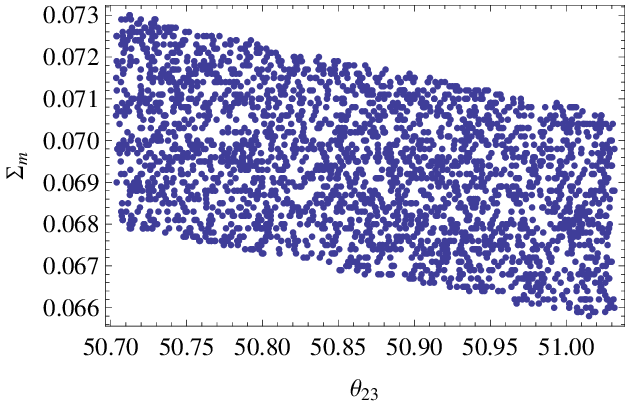}
   \caption{$\Sigma_m$ and $\theta_{23}$ correlation for the class A$_1$ with X$_2$ CP symmetry}   
   \end{subfigure}
\begin{subfigure}{0.4\textwidth}
    \includegraphics[width=\textwidth]{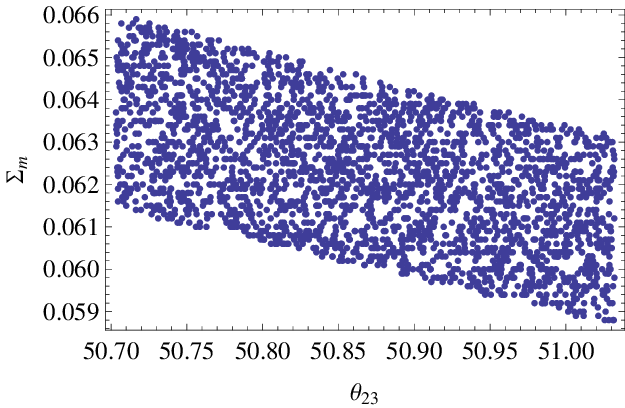}
    \caption{$\Sigma_m$ and $\theta_{23}$ correlation for the class A$_2$ with X$_2$ CP symmetry}   
    \end{subfigure}
 \caption{Correlation plots for the classes A$_1$ and A$_2$ with X$_2$ CP symmetry. All the mixing angles are in degrees and $\Sigma_m$ is in eV.}
\label{figX2A1A2}
\end{figure}

The two patterns $M^{A_1}_{X_2}$ and $M^{A_2}_{X_2}$ of two texture zeros with $X_2$ CP symmetry are consistent with the present neutrino oscillation data. The expression of $R_{\nu}$ has very simple form for both of these patterns and is given by
\begin{equation}\label{rnua1}
{R_{\nu}}^{A_1,A_2}_{X_2} = \sin ^2\theta.
\end{equation}
The allowed parameter space for these patterns  is very restricted. Since, the mass matrix is real and can be diagonalized with TM$_2$ mixing, the mixing angles in terms of the parameter $\theta$ are given by
\begin{equation}\label{eq:th12}
s_{12}^{2} = \frac{1}{3-2 \sin^2\theta},
\end{equation}
\begin{equation}\label{eq:th23}
s_{23}^{2}=\frac{1}{2} \left(1+\frac{\sqrt{3}   \sin 2 \theta}{3-2 \sin^2\theta}\right),
\end{equation}
and
\begin{equation}\label{eq:th13}
s_{13}^{2}=\frac{2}{3}\sin^2\theta.
\end{equation}
Using Eq. \ref{eq:th13} to rewrite the expression of $R_{\nu}$ (given in Eq. \ref{rnua1}) in terms of $\theta_{13}$, we get
\begin{equation}
{R_{\nu}}^{A_1,A_2}_{X_2} = \frac{3}{2}\sin ^2\theta_{13}.
\end{equation}
The mixing angle $\theta_{23}$ takes its extreme value in the higher octant while the CP-violating phase $\delta= 0$.  The Majorana phases can have values $0$, $\frac{\pi}{2}$, and $\pi$. The correlation plots for various parameters of these patterns are shown in Fig. \ref{figX2A1A2}. We observe  that the correlation plot between $\theta_{12}$ and $\theta_{23}$ is identical for both the patterns A$_1$ and A$_2$.

\subsection{One texture zero with one generalized CP symmetry}

There are 12 patterns of neutrino mass matrix with one texture zero and one generalized CP symmetry (6 possibilities for texture zeros and 2 for generalized CP symmetry) as described in the last section. All these patterns will have distinct predictions for neutrino masses, mixing angles, and CP violation.

The neutrino mixing matrix $U$ for any of these patterns depends upon the free parameters which are three angles ($\theta_1$, $\theta_2$, and $\theta_3$) defining the orthogonal matrix $O_{\nu}$ (Eq. \ref{eq:onu}) and the choices of the integers ($k_1$, $k_2$, and $k_3$) defining the matrix $Q_{\nu}$ (Eq. \ref{eq:qnu}). For each pattern of the neutrino mass matrix, we can calculate the parameter $R_{\nu}$, the three neutrino mixing angles, and the three CP-violating phases as described in section \ref{sec:2.4}. 

The angles $\theta_1$, $\theta_2$, and $\theta_3$ defining the generalized CP transformations X$_1$ and X$_2$ are the three free parameters in our numerical analysis of the neutrino mass matrices with one texture zero and a generalized CP symmetry.  We vary these three angles uniformly in the interval $[-\frac{\pi}{2},\frac{\pi}{2}]$. We  consider all the signs resulting from the choices of the integers $k_1$, $k_2$, and $k_3$. Then, we calculate the values of $R_{\nu}$ for these patterns of the neutrino mass matrix using the relations derived in section \ref{sec:2.4}. Next, we obtain the allowed parameter space for the angles ($\theta_1$, $\theta_2$, and $\theta_3$) for which $R_{\nu}$ lies between its experimentally allowed range at 3 standard deviations. Finally, we calculate the predictions for the three mixing angles and CP-violating phases as explained in Section \ref{secgcp}. By varying the smallest neutrino mass $m_{\circ}$ ($m_1$ for NH and $m_3$ for IH) and the two mass-squared differences ($\Delta m_{12}^2$ and $\Delta m_{13}^2$) in their experimentally allowed ranges at three standard deviations (Table \ref{tabdat}), we also calculate the predictions for the effective neutrino mass ($|M_{ee}|$) measurable in neutrinoless double beta decay experiments and the sum of neutrino masses ($\Sigma_m$) observable in the cosmological experiments. The results of this numerical analysis are presented below for one texture zeros with one generalized CP symmetry.

All the 12 patterns with one texture zero and one CP symmetry are allowed for NH.  However, one of the patterns (texture zero at $M_{11}$) is ruled out for inverted hierarchy because this condition predicts a too large value of $\theta_{13}$ for the IH case. The presence of CP symmetries in the mass matrix doesn't change this property of $M_{11}=0$ already known in the previous studies \cite{Dev:2006if, Merle:2006du, Lashin:2011dn}. So, only 10 patterns with one texture zero and one CP symmetry are allowed for IH.

\begin{figure}
\centering
\begin{subfigure}{0.8\textwidth}
    \includegraphics[width=\textwidth]{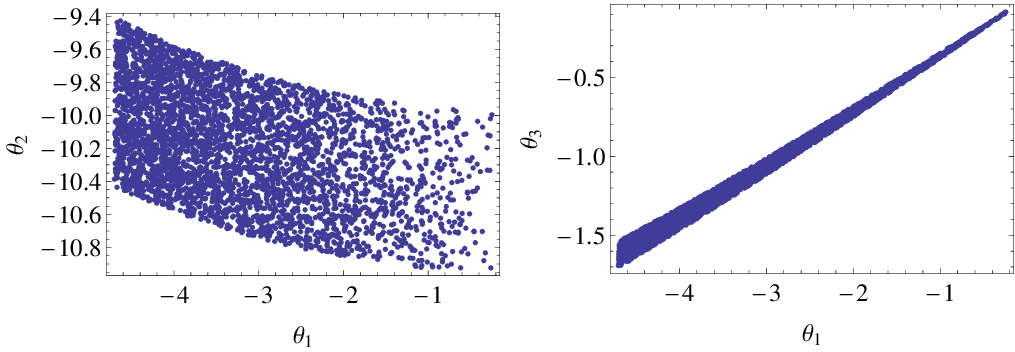}
    \caption{The texture $M_{22}=0$ for X$_1$ CP symmetry in NH}   
    \end{subfigure}
\begin{subfigure}{0.8\textwidth}
    \includegraphics[width=\textwidth]{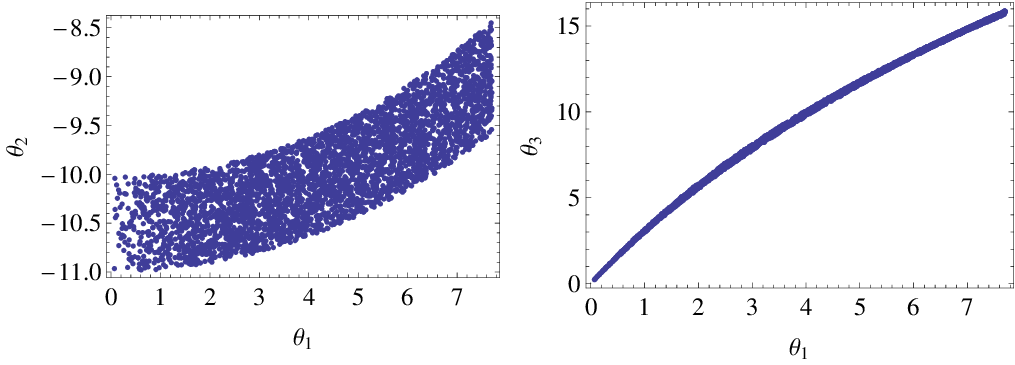}
    \caption{The texture $M_{12}=0$ for X$_1$ CP symmetry  in IH}   
\end{subfigure}
\begin{subfigure}{0.8\textwidth}
    \includegraphics[width=\textwidth]{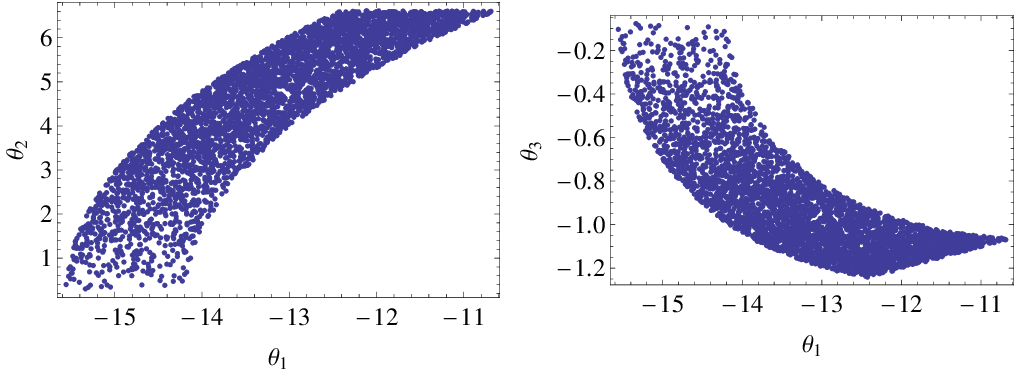}
    \caption{$M_{22}=0$ for X$_2$ CP symmetry  in NH}   
    \end{subfigure}
\begin{subfigure}{0.8\textwidth}
    \includegraphics[width=\textwidth]{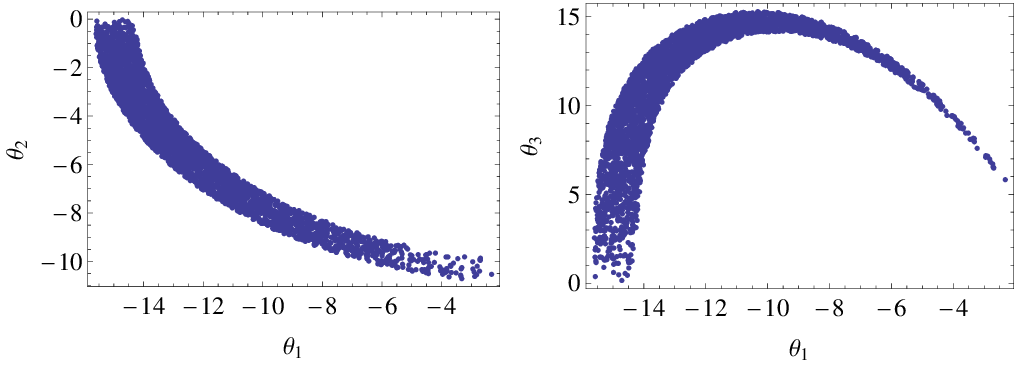}
    \caption{$M_{12}=0$ for X$_2$ CP symmetry  in IH}   
\end{subfigure} 
 \caption{Correlation plots of the free parameters ($\theta_1$, $\theta_2$, and $\theta_3$) for one texture zero with X$_1$/X$_2$ CP symmetry in NH/IH. All angles are in degrees.}
\label{fig:FigPar}
\end{figure}

We find that the allowed values for angles $\theta_1$, $\theta_2$, and $\theta_3$ lie within a narrow range of approximately $[-15^{\circ}, 15^{\circ}]$. In certain patterns, we also found a sharp correlation between $\theta_{1}$ and $\theta_{3}$. The allowed parameter space depends upon the choice of CP symmetry, hierarchy, and position of zero in the neutrino mass matrix.
Figure \ref{fig:FigPar}, depicts the details of the allowed parameter space of angles $\theta_1$, $\theta_2$, and $\theta_3$ for some of these allowed patterns.  

A common feature of various modifications of TBM mixing like TM$_1$ and TM$_2$ is that they predict a neat correlation between $\theta_{23}$ and $\delta$ and as $\theta_{23}$ tends to maximal mixing, CP violation also tends to be maximal ($\delta$ tends to $\frac{\pi}{2}$ or $\frac{3 \pi}{2}$). We obtain similar correlations between $\theta_{23}$ and $\delta$ for CP symmetries X$_1$ and X$_2$ when we impose the texture zeros. The correlations are distinct for each pattern and would act as one of the distinguishing features of these patterns as experiments improve their precision on these parameters.

\begin{figure}
\centering
\begin{subfigure}{0.8\textwidth}
    \includegraphics[width=\textwidth]{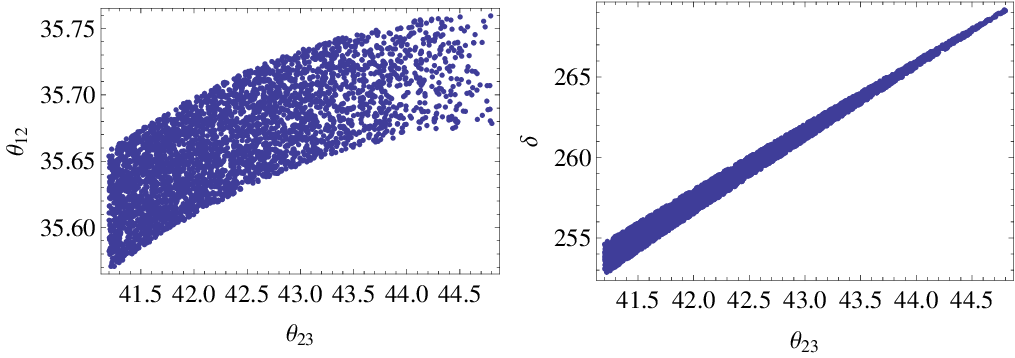}
    \caption{$M_{22}=0$ for X$_1$ CP symmetry  in NH}   
    \end{subfigure}
\begin{subfigure}{0.8\textwidth}
    \includegraphics[width=\textwidth]{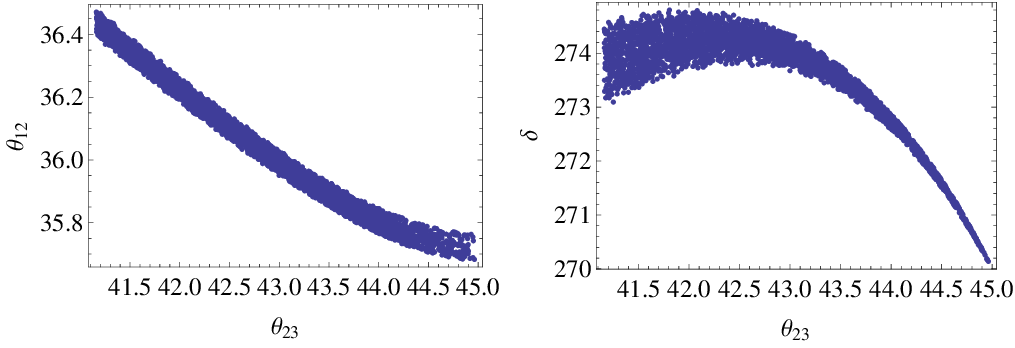}
    \caption{$M_{13}=0$ for X$_1$ CP symmetry  in IH}   
   \end{subfigure}
\begin{subfigure}{0.8\textwidth}
    \includegraphics[width=\textwidth]{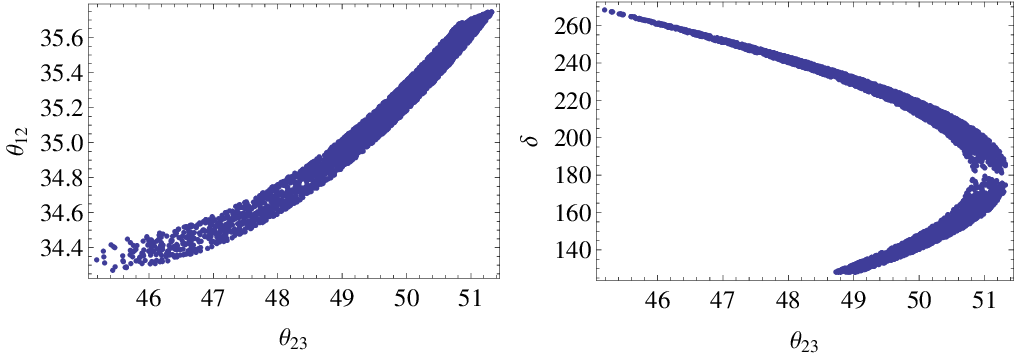}
    \caption{$M_{33}=0$ for X$_2$ CP symmetry  in NH}   
\end{subfigure}
\begin{subfigure}{0.8\textwidth}
    \includegraphics[width=\textwidth]{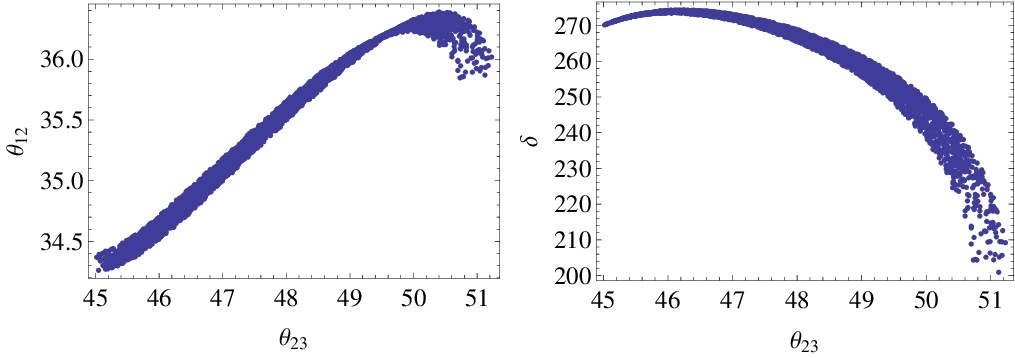}
    \caption{$M_{13}=0$ for X$_2$ CP symmetry  in IH}   
   \end{subfigure}     
 \caption{Correlation plots of Predicted values of the mixing angles ($\theta_{12}$, $\theta_{23}$) and CP-violating phase $\delta$ for one texture zero with X$_1$/X$_2$ CP symmetry in NH/IH.  All angles and phases are in degrees.}
\label{fig:FigAng}
\end{figure}

An interesting prediction of these patterns that distinguishes them from the patterns derived from TM$_1$ and TM$_2$ mixing is a correlation between $\theta_{12}$ and $\theta_{23}$ (and hence with $\delta$).  Clearly, the neutrino mass matrices with generalized CP symmetries and texture zeros are highly testable and will face a close scrutiny from the precision measurements. These predictions for correlations between $\theta_{12}$, $\theta_{23}$, and $\delta$ for some of the allowed cases are depicted in Figure \ref{fig:FigAng}.

\begin{figure}
\centering
\begin{subfigure}{0.8\textwidth}
    \includegraphics[width=\textwidth]{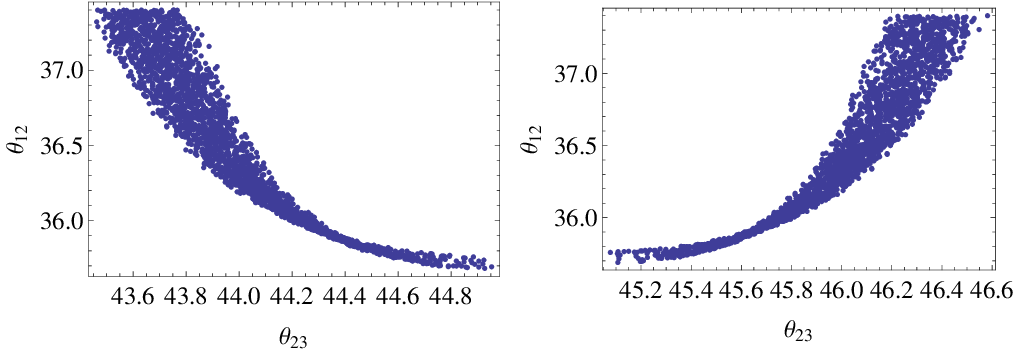}
    \caption{$M_{12}=0$ (Left) and $M_{13}=0$ (Right) for X$_1$ CP symmetry  in NH}   
    \end{subfigure}
\begin{subfigure}{0.8\textwidth}
    \includegraphics[width=\textwidth]{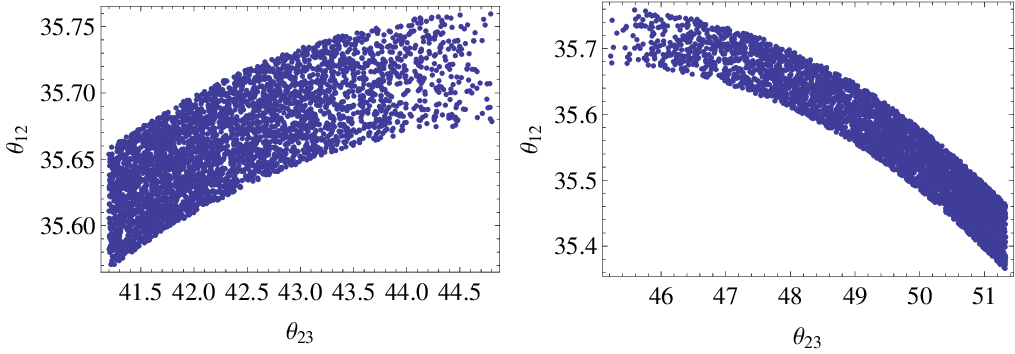}
    \caption{$M_{22}=0$ (Left) and $M_{33}=0$ (Right) for X$_1$ CP symmetry  in NH}   
   \end{subfigure}
 \caption{Correlation plots of Predicted values of the mixing angles $\theta_{12}$ and $\theta_{23}$ for one texture zero with X$_1$ CP symmetry in NH showing complementarity between predictions for $\theta_{23}$. All the angles are in degrees.}
\label{fig:FigComp}
\end{figure}

Like all neutrino mass matrices with texture zeros, with or without additional symmetries, there exists a complementarity between predictions for $\theta_{23}$ and $\delta$ among certain patterns related through 2-3 permutation symmetry \cite{Fritzsch:2011qv}. The pattern $M_{12} = 0$ is complementary to $M_{13} = 0$ and the pattern $M_{22} = 0$ is complementary to $M_{33} = 0$. This complementarity arises because the expression of $M_{12}$ ($M_{22}$) in terms of masses, mixing angles, and phases transforms to $M_{13}$ ($M_{33}$) under exchange
\[
\theta_{23} \rightarrow \frac{\pi}{2}-\theta_{23}~ \text{and}~ \delta \rightarrow \pi + \delta. 
\]
Hence, the predictions for $\theta_{23}$ for the case $M_{13}=0$ ($M_{33}=0$)  is a mirror reflection of the prediction for $M_{12}=0$ ($M_{22}=0$) around $\frac{\pi}{4}$. Also, the prediction for $\delta$ for the case $M_{13}=0$ ($M_{33}=0$)  is a translation of the prediction for $M_{12}=0$ ($M_{22}=0$)  by  a constant angle $\pi$. However, the allowed region for the $\theta_{23}-\delta$ parameter space for the  recent  experimental results is not symmetric under above transformation. In the experimental data, the values of $\theta_{23}$ higher than $\frac{\pi}{4}$ are slightly preferred than the values less than $\frac{\pi}{4}$ and large portion of the values of $\delta$ in the interval $[0,\pi]$ are ruled out in a statistically significant manner. This asymmetry in the data makes our predictions for these parameters asymmetric as well and the complementarity pointed out above will not be evident in the allowed parameter space after imposing the experimental constraints on  $\theta_{23}$ and $\delta$. However, it is still useful in understanding many of the predictions for $\theta_{23}$ and $\delta$.  One interesting example is the prediction that $\theta_{23 }$ is allowed to be below or above maximal mixing for certain patterns. For patterns $M_{12}=0$ and  $M_{22}=0$ with NH for X$_1$ CP symmetry, $\theta_{23}< \frac{\pi}{4}$ and for patterns $M_{13}=0$ and  $M_{33}=0$ with NH for X$_1$ CP symmetry, $\theta_{23}> \frac{\pi}{4}$ as shown in Figure \ref{fig:FigComp}. 

\begin{figure}[t]
\centering
\begin{subfigure}{0.8\textwidth}
    \includegraphics[width=\textwidth]{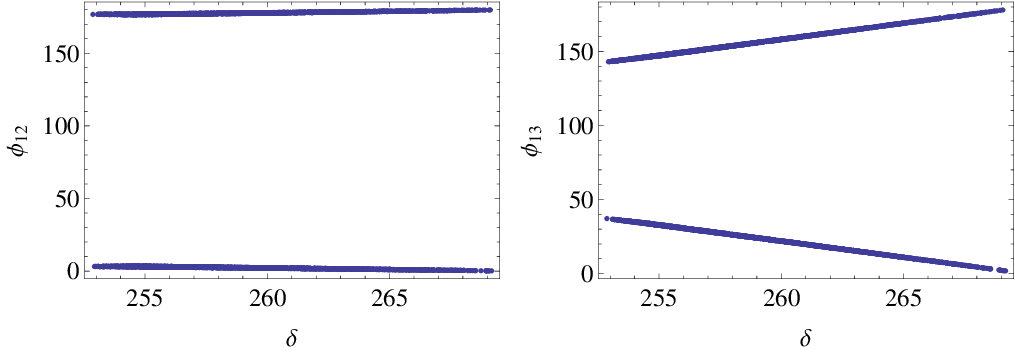}
    \caption{$M_{22}=0$ for X$_1$ CP symmetry  in NH}   
    \end{subfigure}
\begin{subfigure}{0.8\textwidth}
    \includegraphics[width=\textwidth]{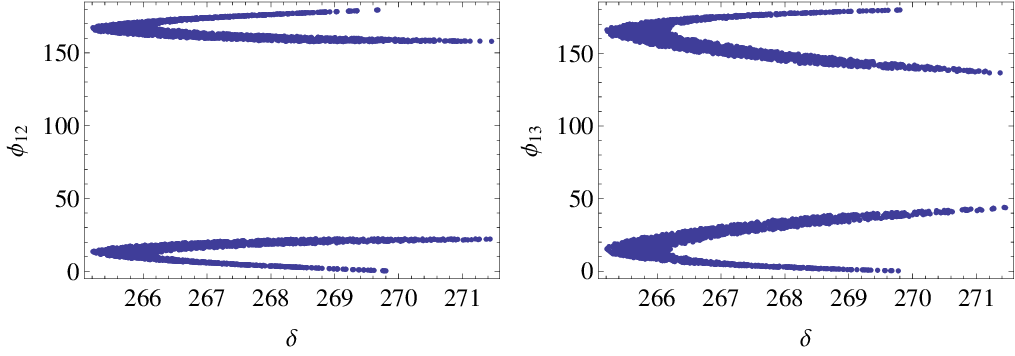}
    \caption{$M_{12}=0$ for X$_1$ CP symmetry  in IH}   
\end{subfigure}
\begin{subfigure}{0.8\textwidth}
    \includegraphics[width=\textwidth]{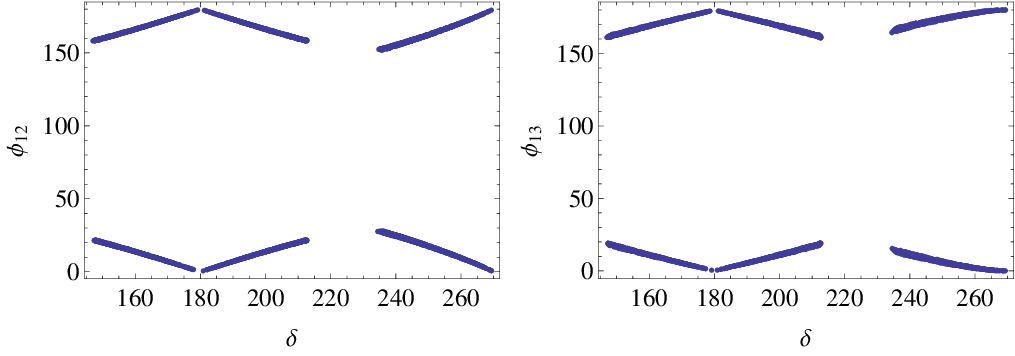}
    \caption{$M_{13}=0$ for X$_2$ CP symmetry  in NH}   
   \end{subfigure}
\begin{subfigure}{0.8\textwidth}
    \includegraphics[width=\textwidth]{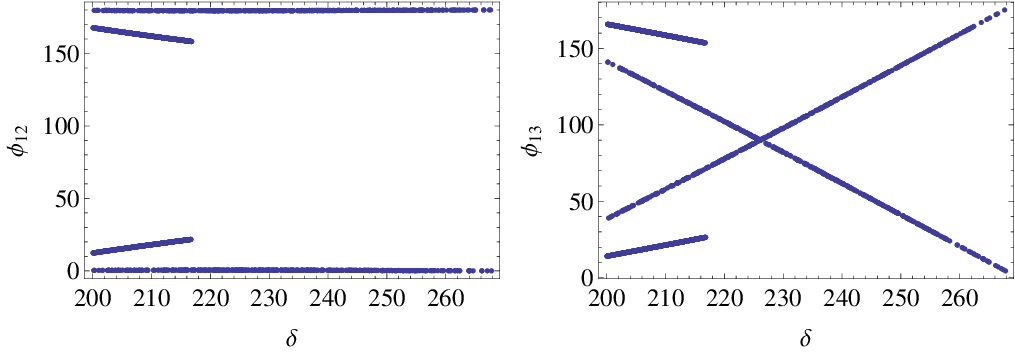}
    \caption{$M_{22}=0$ for X$_2$ CP symmetry  in IH}   
    \end{subfigure}   
 \caption{Correlation plots of the CP-violating phases ($\phi_{12}$, $\phi_{13}$, and $\delta$) for one texture zero with X$_1$/X$_2$ CP symmetry with NH/IH.  All phases are in degrees.}
\label{fig:FigPhases}
\end{figure}

The correlations between the CP-violating phases $\delta$, $\phi_{12}$, and $\phi_{13}$ are almost linear with reflection symmetry around the  $\phi_{12} = \frac{\pi}{2}$ or $\phi_{13} = \frac{\pi}{2}$ lines. In certain cases, the phases $\phi_{12}$ and $\phi_{13}$ take very narrow ranges. The CP-violating phase $\delta$ can act as a discriminator among these various cases. Some of these predictions have been depicted in Figure \ref{fig:FigPhases}.

Cosmological observations have put an upper bound on the sum of neutrino masses ($\Sigma_m = m_1+m_2+m_3 $). Data from Planck satellite combined with baryon acoustic oscillations (BAO) measurements limit the sum of neutrino masses $\Sigma_m \leq 0.12$ eV at 95$\%$ CL \cite{Planck:2018vyg, Tanseri:2022zfe}. However, these limits depend upon certain model specific assumptions. In the present work, we assume a more conservative limit of $\Sigma_m \leq 1$ eV. 

For all of the allowed patterns, we calculate the effective neutrino mass for neutrinoless double beta decay ($|M_{ee}|$) (except for the texture zero at the element $M_{11}$) and the sum of neutrino masses ($\Sigma_m $). The values of $|M_{ee}|$, $\Sigma_m$, and the lowest mass $m_{\circ}$ ($m_1$ for NH and $m_3$ for IH) for all the allowed cases have been summarized in Tables \ref{table3} and \ref{table4}. For the patterns $M_{\mu \mu}=0$, $M_{\mu \tau}=0$, and $M_{\tau \tau}=0$, the predicted values of $\Sigma_m$ are always larger than 0.12 eV for all choices of CP symmetries (X$_1$ and X$_2$) and hierarchies (NH and IH) and tend to violate the Plank data. These patterns will be completely ruled out as the cosmological data becomes more restrictive on $\Sigma_m$ in future. We also note that there are three patterns of neutrino mass matrix with IH for which the lowest neutrino mass $m_3$ can be zero.

\section{Conclusions \label{sec4}}

In this work, we have studied the properties of neutrino mass matrices with texture zeros and CP symmetries. We find that only two cases of two texture zeros are allowed in presence of the X$_1$ or X$_2$ CP symmetries. The neutrino mass matrices in those two patterns are real and predict extreme values of $\theta_{23}$. For neutrino mass matrices with one texture zero in presence of one generalized CP symmetry, all the cases except $M_{11}=0$ for inverted hierarchy are allowed. We find that the angles $\theta_1$, $\theta_2$, and $\theta_3$ fall within a narrow range of approximately $[-15^{\circ}, 15^{\circ}]$ and they are correlated among themselves. We also observe correlations between $\theta_{12}$, $\theta_{23}$, and $\delta$ for CP symmetries X$_1$ and X$_2$, which are distinct for each texture and could be used as their distinguishing features. We note that the complementarity between predictions for $\theta_{23}$ and $\delta$ for certain pairs of texture zeros does not hold exactly due to recent experimental results. We also calculate the effective neutrino mass for neutrinoless double beta decay and the sum of the neutrino masses for all the relevant cases. Certain patterns of one texture zero and one CP symmetry are currently disfavored by the cosmological data and might be completely ruled out as the cosmological data becomes more restrictive in the future. Our results demonstrate the predictability and testability of neutrino mass matrices with one CP symmetry and texture zeros. Their predictions for correlations between mixing angles and CP-violating phases can be used to distinguish these textures from one another as well as from the other texture zero schemes proposed in the literature.

\begin{table*}[t]
\begin{center}
\begin{tabular}{|ccccccc|}
 \hline
 & & Normal Ordering& & & Inverted Ordering & \\
\hline
  & $|M_{ee}|$ (eV) & $m_\circ$ (eV)& $\Sigma_m$ (eV)& $|M_{ee}|$ (eV)& $m_\circ$ (eV)& $\Sigma_m$ (eV)\\
 \hline 
$M_{ee} = 0$ & 0 & 0.003-0.004  & 0.058 - 0.077 & - & - & - \\
 &  & $\oplus$ 0.005- 0.008 &  &  &  &  \\
$M_{e \mu} = 0$ &  0.024 - 0.221 & 0.023 - 0.221 & 0.099 - 0.670 &  0.057 - 0.277 & 0 - 0.272 & 0.099 - 0.830 \\
$M_{e \tau} = 0$ &  0.0233 - 0.174 & 0.023 - 0.017 & 0.098 - 0.525 &  0.049 - 0.313 & 0.009 - 0.310 & 0.109 - 0.933  \\
$M_{\mu \mu} = 0$ & 0.054 - 0.310 & 0.054 - 0.310 & 0.181 - 0.933 &   0.063 - 0.331 & 0.041 - 0.330 & 0.172 - 1.000  \\
$M_{\mu \tau} = 0$ & 0.139 - 0.209 & 0.145 - 0.215 & 0.445 - 0.652  &  0.016 - 0.017 & 0.017 - 0.018 & 0.120 - 0.125 \\
$M_{\tau \tau} = 0$ & 0.037 - 0.284 & 0.037 - 0.283 & 0.135 - 0.860 & 0.076 - 0.320 & 0.058 - 0.316 & 0.210 - 0.957   \\
 \hline 
 \end{tabular}
\caption{Allowed ranges of $|M_{ee}|$, $m_\circ$, $\Sigma_m$ for X$_1$ CP symmetry.}
\label{table3}
\end{center}
\end{table*}

\begin{table*}[t]
\small
\begin{center}
\begin{tabular}{|ccccccc|}
 \hline
 & & Normal Ordering& & & Inverted Ordering & \\
\hline
  & $|M_{ee}|$ (eV) & $m_\circ$ (eV)& $\Sigma_m$ (eV)& $|M_{ee}|$ (eV)& $m_\circ$ (eV)& $\Sigma_m$ (eV)\\
\hline 
$M_{ee} = 0$ & 0 & 0.0021 - 0.0085 & 0.058 - 0.077 & - & -  & -\\
$M_{e\mu} = 0$ & 0.0010 - 0.0014  & 0.0067 - 0.0100  & 0.060 - 0.085  & 0.047 - 0.317 & 0 - 0.313 & 0.098 - 0.954\\
& $\oplus$ 0.018 - 0.312 &  $\oplus$ 0.018 - 0.300 & $\oplus$ 0.095 - 0.940 &    &   & \\
$M_{e\tau} = 0$ &  0.0001 - 0.0018 & 0.0026 - 0.0046 & 0.033 - 0.074 & 0.047 - 0.314 & 0 - 0.311  & 0.098 - 0.940\\
 & $\oplus$ 0.019 - 0.200 & $\oplus$ 0.018 - 0.200 & $\oplus$ 0.091 - 0.600 &  &  &  \\
$M_{\mu \mu} = 0$ & 0.05 - 0.32  & 0.05 - 0.32 & 0.17 - 0.96 & 0.0172 - 0.0179 & 0.023 - 0.025 & 0.132 - 0.138  \\
&   &  &  & $\oplus$ 0.066 - 0.270 & $\oplus$ 0.047 - 0.265 & $\oplus$ 0.18 - 0.81 \\
$M_{\mu \tau} = 0$ & 0.137 - 0.328 & 0.144 - 0.333 & 0.441-1 & 0.0158 - 0.0166 & 0.0193 - 0.0204 & 0.124 - 0.133\\
&   &  &  & $\oplus$ 0.0187 - 0.0198 & $\oplus$ 0.0211 - 0.0224 &  \\
$M_{\tau \tau} = 0$ & 0.033-0.324 & 0.034-0.325 & 0.128 - 0.980 &  0.015180 - 0.01593 & 0.00685 - 0.00815 & 0.1056 - 0.1110 \\
&   &  &  & $\oplus$ 0.076 - 0.243 & $\oplus$ 0.060 - 0.238 & $\oplus$ 0.214 - 0.730 \\
\hline 
\end{tabular}
\caption{Allowed ranges of $|M_{ee}|$, $m_\circ$, $\Sigma_m$ for X$_2$ CP symmetry.}
\label{table4}
\end{center}
\end{table*}

\bibliographystyle{elsarticle-num}
\bibliography{cpbibliography}

\end{document}